\documentclass[prl,aps,twocolumn]{revtex4}

\usepackage{latexsym,exscale}
\usepackage{graphicx}
\usepackage{amsmath,amssymb,amsfonts}

\begin{document}

\title{Hadronization in heavy ion collisions: 
       Recombination and fragmentation of partons}
\author{R.~J.~Fries}
\affiliation{Department of Physics, Duke University, Durham, NC 27708}
\author{S.~A.~Bass}
\affiliation{Department of Physics, Duke University, Durham, NC 27708}
\affiliation{RIKEN BNL Research Center, Brookhaven National Laboratory, 
        Upton, NY 11973, USA}
\author{B.~M\"uller}
\affiliation{Department of Physics, Duke University, Durham, NC 27708}
\author{C.~Nonaka}
\affiliation{Department of Physics, Duke University, Durham, NC 27708}
\date{\today}

\begin{abstract}
We argue that the emission of hadrons with transverse momentum up to
about 5 GeV/$c$ in central relativistic heavy ion collisions is 
dominated by recombination, rather than fragmentation of partons.
This mechanism provides a natural explanation for the observed 
constant baryon-to-meson ratio of about one and the apparent lack 
of a nuclear suppression of the baryon yield in this momentum range.
Fragmentation becomes dominant at higher transverse momentum, but
the transition point is delayed by the energy loss of fast partons 
in dense matter.
\end{abstract}

\maketitle

Data from the Relativistic Heavy Ion Collider (RHIC) have shown a strong
nuclear suppression of the pion yield at transverse momenta larger than
2 GeV/$c$ in central Au + Au collisions, compared with $p+p$ interactions 
\cite{PHENIX}. The emission of
protons and antiprotons does not appear to be similarly suppressed, and 
the $p$/$\pi^+$ ratio reaches or even exceeds unity for transverse 
momenta above 2 GeV/$c$ \cite{PHENIX-B,STAR-B,STAR-L,PHENIX-L}. These 
results lack a consistent explanation in the standard picture of hadron 
production at high transverse momentum, which assumes that hadrons 
are created by the fragmentation of energetic partons. 
Whereas the observed suppression of the pion yield is attributed 
to the energy loss of partons during their 
propagation through the hot and dense matter created in the nuclear 
collision -- a phenomenon commonly referred to as jet quenching 
\cite{GyulWang:94} -- the absence of a similar effect in the proton 
spectrum is puzzling \cite{Science}. 

We propose that hadron production at momenta of a few GeV/$c$ in an 
environment with a high density of partons occurs by recombination, 
rather than fragmentation, of partons. Below we show that recombination 
always dominates over fragmentation for an exponentially falling parton 
spectrum, but that fragmentation wins out eventually, when the spectrum 
takes the form of a power law. We also show that recombination can 
explain some of the surprising features of the RHIC data, as first suggested
by Voloshin \cite{Voloshin:02}.

In the fragmentation picture \cite{CoSo:81} the single 
parton spectrum is convoluted with the probability for a
parton $i$ to hadronize into a hadron $h$, which carries a fraction 
$z<1$ of the momentum of the parent parton.  It has been argued that 
the fragmentation functions $D_{i\to h}(z)$ can be altered by 
the environment \cite{GW:00}. The dominant modification 
mechanism is the energy loss of the propagating parton in the surrounding 
medium, which leads, in first approximation, to a rescaling of the 
variable $z$. This would affect all produced hadrons in the same way, 
in contradiction with the observations at RHIC.

Another mechanism of hadron production is quark recombination. 
In the recombination picture, three  quarks or a quark/antiquark pair 
in a densely populated phase space
can form a baryon or meson respectively. The amplitude for this process
is determined by the hadron wave function. This mechanism 
has recently been identified as the source of unnatural isospin ratios 
in the production of $D$-mesons in the fragmentation region in $\pi^- +A$ 
interactions at Fermilab \cite{BJM:02}. Hadron production in heavy ion 
collisions by recombination of quarks has been considered before 
\cite{recomb}, primarily at small transverse momentum.  
Quark recombination has recently been 
invoked to explain some aspects of the RHIC data, such as the flavor 
pattern of elliptic flow \cite{LinKo:02}, and in the context of a 
scaling model \cite{HwaYa:02}.

In the formalism of perturbative QCD, recombination is a more exclusive
process, and falls off faster than fragmentation with increasing transverse 
momentum. On the other hand, for a meson of given momentum $P$ fragmentation 
starts out with a parton with much higher momentum $p=P/z$, $(z<1)$, 
whereas recombination requires a quark/antiquark pair where each parton 
carries only about $P/2$ in average. However, the spectrum of high-momentum 
partons is steeply falling and further reduced by the energy loss in dense 
matter.
Our main result is that the momentum range, in which recombination processes 
can successfully compete with fragmentation, may extend up to 5 GeV/$c$ in the 
favorable environment of central heavy ion collisions at RHIC.
We will show below that recombination is effective whenever the phase-space
distribution of a system of partons has thermal character.

Let us consider an expanding system of quarks and antiquarks. We assume 
that the recombination of these partons into hadrons occurs on a space-like 
hypersurface $\Sigma_{\text{f}}$. The RHIC experiments indicate that the 
freeze-out is very rapid. The measured two-particle correlation functions 
are consistent with an extremely short emission time in the local rest 
frame, suggesting a sudden transition after which individual hadrons 
interact only rarely \cite{RHIC-HBT}. In our treatment, 
we assume that the dense parton matter is devoid of dynamical thermal 
gluons and predominantly composed of quarks and antiquarks at the moment 
of hadronization. We neglect a possible effective quark mass, because we 
are here interested in hadron production at high transverse momentum.

Denoting the density matrix of the parton system on $\Sigma_{\text{f}}$ as 
$\hat\rho_{\text{f}}$, we can express the number of mesons at freeze-out as
$  E {dN_M}/{d^3P} =  (2\pi)^{-3} \> 
  \langle M;{ P} | \> \hat\rho_{\text{f}} \> | M;{ P} \rangle \,$ .
Here $| M;{ P} \rangle$ is a meson state with momentum $P$. 
We now introduce the single-particle Wigner functions for quarks and 
antiquarks, $w_a(r;p)$ and $\bar w_b(r;p)$, respectively, and neglect 
multi-particle correlations in the density matrix.
Further introducing the time-like, future oriented unit vector $u^\mu(r)$ 
orthogonal to the freeze-out hypersurface at $r \in \Sigma_{\text{f}}$ 
\cite{CoFr:74,DHSZ:91}, we obtain the following expression for the
meson emission spectrum:
\begin{widetext}
\begin{equation}
  \label{eq:mesf}
  E \frac{d N_M}{d^3 P} = 
  \int\limits_{\Sigma_{\text{f}}} d\sigma \, \frac{P\cdot u(r)}{(2\pi)^3} \,
 \sum_{a,b} \int\limits_0^1 dz \, | \psi^{M}_{ab}(z) |^2 
  w_a(r;zP^+) \, \bar w_b(r;(1-z)P^+) \,.
\end{equation} 
Because we are considering the emission of a meson with high momentum, 
it is most convenient to use a distribution amplitude in terms of light-cone 
coordinates in a frame where $P^+ \gg P^-$ and the transverse momentum 
of the meson vanishes. $z$ and $1-z$ are the momentum fractions carried 
by the quark and antiquark, respectively, and $a,b$ denote the internal 
quantum numbers of the quarks (spin, flavor, color).  
For baryons, a similar calculation yields:
\begin{equation}
  \label{eq:barf}
  E \frac{d N_B}{d^3 P} = 
  \int\limits_{\Sigma_{\text{f}}} d\sigma \, \frac{P\cdot u(r)}{(2\pi)^3} \, 
  \sum_{a,b,c} \int\limits_0^1 dz_1 \int\limits_0^{1-z_1} dz_2 \, 
  | \psi^{B}_{abc}(z_1,z_2) |^2 
  w_a(r;z_1P^+) \, w_b(r;z_2P^+) \, w_c(r;(1-z_1-z_2)P^+) \, .
  \end{equation}
\end{widetext}
The overall normalization is fixed by the number of quarks on the 
hypersurface $\Sigma_{\text{f}}$:
\begin{equation}
  \label{eq:norm} 
  N_a = \int_{\Sigma_{\text{f}}} d\sigma \frac{d^3 p}{(2\pi)^3 p^0} 
  \, p\cdot u(r) w_a(r;p) \, .
\end{equation}

It is important to note that our results are not really sensitive to the 
model used for the recombination process. A complete dynamical 
description in terms of QCD is hard to achieve. However, for the observables 
that we discuss below, it is essential to use two common features: 
the probability for the emission of a meson (baryon) is proportional to 
the single parton distribution squared (cubed), and the parton 
momenta sum up to the hadron momentum. At this level of sophistication 
a description with the light-cone wave functions replaced by 
three-dimensional, spatial quark wave functions leads to identical results.

To make further progress, we need to specify the parton Wigner functions.
We first consider the case that these are described by the exponential 
tail of a thermal distribution $w_a(r;p) = \exp(- (p\cdot u - \mu)/T)$ 
with local temperature $T(r)$ and chemical potential $\mu(r)$, 
independent of the internal quantum numbers. The constraint that
the momentum fractions of all partons in the hadron wave function must
add up to the total momentum $P$ of the hadron then insures that the
product of all Wigner functions entering into the hadron yield is
solely dependent on the hadron momentum:
\begin{align}
  \label{eq:therm}
  w_a(r;zP^+) \, \bar w_b(r;(1-z)P^+) =  \exp(- P\cdot u/T), &
  \nonumber \\
  w_a(r;z_1P^+) \, w_b(r;z_2P^+) \, w_c(r;(1-z_1-z_2)P^+) 
  &   \nonumber \\ = \exp(- (P\cdot u - \mu_B)/T) \, & ,
\end{align}
where $\mu_B = 3\mu$. One can then perform the integrations over the
momentum fractions in (\ref{eq:mesf}) and (\ref{eq:barf}) and obtains 
the result that the baryon-to-meson ratio is independent of the momentum
$P$ and simply given by the ratio of the number of quark degrees of
freedom contributing to the emission of the specific hadrons:
${dN_B}/{dN_M} = {\sum_{a,b,c} e^{\mu_B/T}}/{\sum_{a,b}}$.

The summations over color, flavor, and helicity give rise to degeneracy 
factors for mesons and baryons, $C_M$ and $C_B$, respectively. 
In order to derive an upper limit to the nucleon-to-pion ratio,
we make the simplifying assumption that all decuplet baryon states 
$\Delta$ contribute to the nucleon yield through decay, but neglect 
the contributions to the high-$P_T$ pion spectrum from the decay of
unstable hadrons. If we require the partons to form color singlets 
at recombination, we obtain $C_{\pi^+} = 1$ and 
$C_p = 20/(2\times 3!) = 5/3$, yielding the result
\begin{equation}
  \label{eq:ptopi}
  dN_p/dN_{\pi^+} = e^{\mu_B/T} C_p / C_{\pi^+} = 
  {5\over 3} \, e^{\mu_B/T} \, .
\end{equation}
Decay of unstable hadron states is likely to reduce this value, but 
will not change the prediction that the $p/\pi^+$ ratio is constant as 
a function of particle momentum and significantly larger than the
value expected from quark fragmentation. We note that this result
is in good agreement with the RHIC data, which show $dN_p/dN_{\pi^+}$
reaching a plateau around unity in the range 2 GeV/$c$ $\le P_T \le
4$ GeV/$c$ \cite{RHIC-ppi}. For the $\Lambda$-hyperon yield, we need to
include the channels $\Lambda$, $\Sigma^0$ and $\Sigma^{*0}$, leading to
$C_\Lambda = 4/3$. This would
predict a plateau at 4/3 in the $\Lambda/K^0_s$
yield, which will be diluted by kaons from
the decay of $K^*$. The $\bar p$/$p$ ratio is also predicted
to be independent of momentum and equal to $\exp(-2\mu_B/T)$, again
in rather good agreement with the RHIC data \cite{STAR-B}.

The predictions are radically different, when one considers a power law 
spectrum as it is characteristic in perturbative QCD at large transverse 
momentum: 
\begin{equation}
  w_a^{\text{pert}}(r;p) = A_a \left(1+\frac{p_T}{B}\right)^{-\alpha}.
\end{equation}
One then finds that mesons always dominate over baryons at large momentum,
$dN_B/dN_M = (27/4P)^\alpha C_B A_q^2 / C_M A_{\bar q}$,
and that eventually parton fragmentation wins out over quark recombination. 
For pions, the local ratio of the two contributions is
${dN_\pi^{\rm frag}}/{dN_\pi^{\rm rec}} \propto P^{\alpha}$.
On the other hand, for an exponential quark spectrum, fragmentation 
is always suppressed with respect to recombination. 

This result
constitutes the main insight gained from our considerations: Hadron
emission from a thermal parton ensemble is {\em always} dominated
by parton recombination; only when the thermal distribution gives
way to a perturbative power law at high momentum, does fragmentation
become the leading hadronization mechanism. The threshold between
the two domains depends on the size of the emitting system and the hadron
species. 

That the meson spectrum from recombination is 
determined, on the average, by $w(P/2)$, whereas the baryon spectrum 
depends on $w(P/3)$, implies that those kinematic properties of the 
hadron spectrum, which are due to collective flow of partons, extend to higher 
values of the transverse momentum for baryons than for mesons \cite{LinKo:02}. 
This effect is clearly visible in the RHIC data \cite{RHIC-ppi,RHIC-v2}, 
which exhibit a linear rise of the elliptic flow velocity $v_2$ with $P_T$, 
which continues further in $P_T$ for protons and hyperons than for 
pions and kaons.

\begin{figure}[tb]   
\begin{center}
\includegraphics[width=0.87\linewidth]{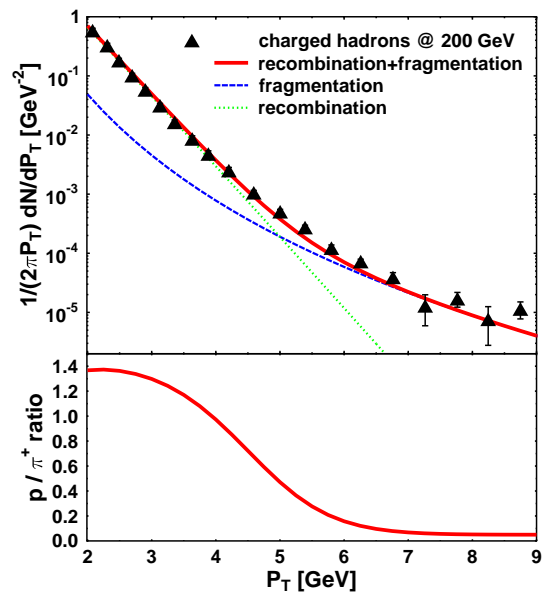}
\end{center}
\caption{Top: inclusive $P_T$ spectrum of charged hadrons in central Au+Au
collisions at $\sqrt{s_{\text{NN}}}=200$ GeV; data taken from the PHENIX
collaboration. Bottom: Ratio of protons to $\pi^+$ as a function of $P_T$. 
The region below 4 GeV/$c$ is dominated by recombination, the region 
above 6 GeV/$c$ by parton fragmentation. }
\label{FBMN:fig1}
\end{figure}

Realistic calculations require the specification of the freeze-out 
hypersurface $\Sigma_{\text{f}}$ and the parton  spectrum.
Here we assume boost invariance of $\Sigma_{\text{f}}$ according to
$\tau_{\text{f}} = \sqrt{t_{\text{f}}^2 - z_{\text{f}}^2} =$ const. 
\cite{DHSZ:91}. For the thermal part of the quark spectrum we use an 
axially and longitudinally expanding thermal source:
\begin{equation}
  \label{eq:thermal}
  w_a^{\text{th}} (\eta;y,p_T) = 
  A_{\text{th}} e^{-p_T\cosh(\eta-y)/T} e^{-y^2/2\Delta^2},
\end{equation}
characterized by an effective temperature $T\approx 350$ MeV, which
includes the blue shift caused by the radial expansion, and a rapidity 
width $\Delta\approx 2$. Here $\eta$ and $y$ are the rapidities in space-time 
and momentum. The form of this spectrum agrees with the
results of the parton cascade {\sc VNI/BMS} \cite{PCM}, which yields a parton
distribution exhibiting an exponential shape at low transverse momentum 
and a power law shape for high transverse momentum. For the power law 
tail of the parton spectrum we choose the results given by a lowest-order
perturbative QCD calculation \cite{FMS:02}, shifted by 
$\Delta p_T = -\sqrt{\lambda p_T}$ with $\lambda=1\text{ GeV}$
to account for the energy loss of fast partons, and standard
fragmentation functions \cite{KKP:00}. The normalization
of the thermal part of the spectrum is adjusted to fit the measured
inclusive spectrum of charged hadrons from PHENIX \cite{Jia:02},
as shown in the upper frame of Fig.~\ref{FBMN:fig1}.
The contributions from recombination and fragmentation are shown separately 
to exhibit the location of the rapid crossover between these two
mechanisms at about 5 GeV/$c$. We note that in the parton spectrum 
$w^{\text{th}}+w^{\text{pert}}$ itself, the crossover between the 
thermal and the perturbative part occurs already at about 3 GeV/$c$, 
consistent with parton cascade predictions \cite{PCM}.
The recombination mechanism shifts this point to higher values of 
$p_T$ in the hadron spectrum. 

The lower frame of Fig~\ref{FBMN:fig1} shows our prediction 
for the $p/\pi^+$ ratio. The rapid drop of its value in the range 
4--5 GeV/$c$ is an unambiguous prediction of our model. Experiments
at RHIC have not yet been able to probe this $P_T$ range, because 
the identification of protons has not been feasible beyond 4 GeV/$c$.
The identification of hyperons is possible to higher $P_T$,
and indications of a rapid drop in the $\Lambda/K^0_s$ ratio have
been found \cite{STAR-L}. 

\begin{figure}[tb]   
\begin{center}
\includegraphics[width=0.86\linewidth]{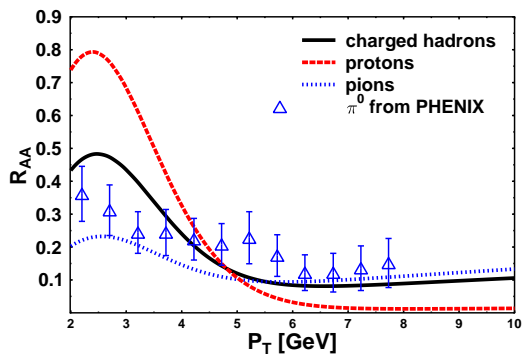}
\end{center}
\caption{Ratio $R_{\text{AA}}$ of hadron yields in central Au+Au 
to $p+p$ collisions, scaled by the number of binary nucleon-nucleon
interactions. Data for pions taken from PHENIX \cite{Miod:02}.}
\label{FBMN:fig2}
\end{figure}

Figure \ref{FBMN:fig2} shows the scaled ratio of particle yields in 
Au + Au and $p+p$ collisions, called $R_{\text{AA}}$. (We use fits to
data for the $p+p$ yields and our predictions for Au + Au.) The energy 
loss of fast partons leads to a nuclear suppression in the fragmentation 
region ($P_T > 5$ GeV/$c$). For low $P_T$ this suppression is counteracted
by the recombination mechanism, which is absent in $p+p$ reactions. 
Recombination is more important for protons than for pions, resulting
in much less nuclear suppression for protons.

RHIC data \cite{RHIC-v2} exhibit a strong increase of the anisotropic flow 
parameter $v_2$ for mesons and baryons at small $P_T$, which finally 
saturates. This happens earlier for mesons than for baryons. It has been 
argued that the flow anisotropy originates in the partonic phase 
\cite{Kolb:00}. In the recombination region mesons at transverse momentum 
$P_T$ reflect the properties of partons with an average transverse momentum 
$P_T/2$, while baryons reflect those of partons with $P_T/3$. It follows 
that $v_2$ saturates later for baryons than for mesons \cite{Voloshin:02}. 
The transition
to the fragmentation region would provide such a mechanism, but it
occurs at too high momentum. The observed saturation of $v_2$ must, 
therefore, be due to some other mechanism or require a more realistic
description of the space-time evolution of the system.

In summary, we propose a two component behaviour of hadronic observables in
heavy ion collisions at RHIC. These components include fragmentation of 
high-$p_T$ partons and recombination from a thermal parton distribution. 
The competition between recombination and fragmentation of partons can 
explain several of the surprising features of the published data. 
In particular, the proton excess at intermediate $P_T$, the different 
nuclear suppression observed in pion and proton spectra, and the different 
saturation thresholds in the elliptic flow, are easily explained. We predict
that all baryon spectra will exhibit a rapid transition around 5 GeV/$c$ to 
a region dominated by parton fragmentation. Finally, our scenario requires 
the assumption of a thermalized partonic phase characterized by an 
exponential momentum spectrum. Such a phase may be appropriately called
a quark-gluon plasma.

Note added: We draw attention to the closely related recent work of Greco 
{\it et al.}\ \cite{GreKoLe:03}, who also propose that parton recombination 
can explain the large baryon/meson ratio observed at RHIC.

\begin{acknowledgments}
We thank T.\ Mehen and M.\ Asakawa for very helpful discussions. 
This work was supported by RIKEN, Brookhaven National Laboratory, 
DOE grants DE-FG02-96ER40945 and DE-AC02-98CH10886, and by the 
Alexander von Humboldt Foundation.
\end{acknowledgments}

\end{document}